\documentstyle[aps,preprint]{revtex}
\newcommand{\dy}[2]{\frac{{\textstyle \partial \/} #1}{{\textstyle \partial \/} #2}}
\begin{document}
\begin{titlepage}
\title{Quark gluon plasma as a strongly coupled \\ 
color-Coulombic plasma}      
\author{Vishnu M. Bannur \\
{\it Institute for Plasma Research}, \\  
{\it Bhat, Gandhinagar  382428, India.} }
\maketitle
\begin{abstract} 
	We show that the extensively studied equation of state (EOS) of 
strongly coupled QED plasma fits the recent lattice EOS data of gluon 
plasma remarkably well, with appropriate modifications to take account of 
color degrees of freedom and running coupling constant. Hence we conclude 
that the quark gluon plasma near the critical temperature is a strongly 
coupled color-Coulombic plasma.    
\end{abstract}
\vspace{1cm}

\end{titlepage}
Quark gluon plasma (QGP), the deconfined state of quarks and gluons, is a 
prediction of quantum chromodynamics (QCD). In search of QGP, experiments 
are on at CERN, BNL etc. where heavy ions are accelerated to relativistic 
energies and made to collide. At sufficiently large energy ($>$ few GeV), 
the fireball formed by collision may be QGP 
which then expands, cools and freezes out 
into hadrons. These hadrons, photons, leptons etc. are detected and 
analysed to see whether QGP is formed or not. The expansion of QGP, 
generally described by relativistic hydrodynamics, affects the experimental 
observations. In the description of hydrodynamics, the EOS of QGP is 
needed to complete the set of fluid equations. At present all analysis of 
experimental results are based on ideal equation of state where quarks and 
gluons are noninteracting and pressure or energy density are propotional to 
the $4^{th}$ power of temperature. In the theoretical calculations of 
various signatures of QGP, it is assumed that QGP is an ideal plasma.  

However, the recent lattice simulation results show that gluon plasma 
is not ideal and is nonideal even upto five times the critical temperature 
($T_c$) \cite{bo.1}. There were lot of attempts \cite{ri.1,pe.1,ba.1} to explain 
this numerical experiment 
based on the properties of QCD, confinement and asymptotic freedom, 
properties of plasma, etc. All of them, so far, not able to explain the 
data satisfactorily. 

In Ref. \cite{ri.1}, they tried to fit the lattice results assuming Coulombic 
interaction with running coupling constant, confinement in the form of 
bag constant and low momentum cut-off in the evaluation of partition 
function. In Ref. \cite{pe.1}, gluons are assumed as quasi-particles with mass 
propotional to plasma frequency. Here also a modified, temperature 
dependent, running coupling constant is used to fit the old 
lattice results of Ref. \cite{en.1}. As we will see it 
does not fit the recent more refined data \cite{bo.1}. 
Earlier we \cite{ba.1} had assumed 
Cornel potential interaction \cite{ei.1}, Coulomb + confinement, 
between quark and 
antiquark and used Mayer's cluster expansion method to derive EOS. This EOS 
was used to fit lattice results on gluon plasma with partial  
success. Especially near the critical temperature fitting was not 
good. We suspected that it might be due to the fact that our theory was for 
weakly interacting system and hence valid only in the high temperature 
limit. To explain the lattice results near $T_c$ we need a theory of 
strongly interacting system. In other words, near $T_c$ QGP may be a  
strongly coupled plasma. This is exactly what we find in this letter, 
namely, QGP is a strongly coupled color Coulombic plasma. 

Generally when we say plasma, it is a quasi neutral system of 
charged particles which exhibits collective effects. The so called 
plasma parameter $\Gamma$, which is the ratio of average potential 
energy to average kinetic energy of particles, is assumed to be weak 
($<<1$). Lot of studies in plasma such as  plasma waves, instabilities 
and other collective effects are in this range of $\Gamma$. 
For $\Gamma$ close to 1 and above, it is a strongly coupled plasma (SCP) 
which modifies various properties of plasma \cite{ic.1}. 
We see that QGP is also a strongly coupled plasma near $T_c$. 
Partially analytic and partially numerical calculations of 
SCP do exist in the literature \cite{ic.1}. 
In particular, EOS of SCP is extensively 
studied and parameterized as a function of $\Gamma$. 

Let us now discuss about QGP. We take QGP as a deconfined, 
quasi-color-neutral system of quarks, antiquarks and gluons with 
color Coulombic mutual interactions. It is similar to QED plasma 
apart from few modifications due to color degrees of freedom. It is 
charactorized by plasma parameter, 
\begin{equation} \Gamma \equiv \frac{<PE>}{<KE>} = \frac{ \frac{4}{3} \frac{\alpha_s}{r_{av}} }{T} \; ,\end{equation}
where we have taken the well known Coulombic interactions used in 
hadron spectroscopy \cite{ei.1}. The typical value of $\alpha_s \approx 0.5$, 
$r_{av} \approx 1 fm$ and near the critical temperature, $T_c \approx 
200 \; MeV$, we estimate $\Gamma \approx 2/3$. Hence QGP is a strongly 
coupled plasma. Compared to QED, the fine structure constant, $\alpha$, 
is replaced by $4 \; \alpha_s /3$ in Coulomb interaction term. 
$r_{av}$ may be estimated as $r_{av} = (3/4 \pi n)^{1/3}$   
and hence  
\begin{equation} \Gamma = \left( \frac{4 \pi n}{3} \right) ^{1/3} \frac{4}{3} \frac{\alpha_s}{T} 
\; ,  \end{equation}
where '$n$' is the number density. 
Taking $n \approx a_f T^3$ we get 
\begin{equation} \Gamma \approx \left( \frac{4 \pi a_f}{3} \right) ^{1/3} \frac{4}{3} \alpha_s 
\; , \label{eq:ga} \end{equation}
$a_f$ is a constant which depends on degrees of freedom. 
As we discussed earlier there exists EOS for strongly coupled Coulombic 
system as a function of $\Gamma$. Since QGP is also a SCP with $\Gamma$ given 
by Eq. (\ref{eq:ga}), we modify QED nonrelativistic energy density,
 $ \varepsilon_{QED} = (3/2 + u_{ex} (\Gamma) ) \, n \, T $ 
for strongly coupled relativistic QGP (SCQGP) as 
\begin{equation} \varepsilon = (3 + u_{ex} (\Gamma) ) \, n \, T \;, \end{equation}
where 
\begin{equation} u_{ex} (\Gamma) = \frac{u_{ex}^{Abe} (\Gamma) + 3 \times 10^3 \, \Gamma^{5.7} 
  u_{ex}^{OCP} (\Gamma) }{1 + 3 \times 10^3 \, \Gamma^{5.7} } \; . \end{equation}
Here we have taken same $u_{ex}^{Abe}$ and $u_{ex}^{OCP}$ as used 
in SCP and are given by   
\begin{equation} u_{ex}^{Abe} (\Gamma) = - \frac{\sqrt{3}}{2} \, \Gamma^{3/2} - 3 \, \Gamma^3 
 \left[ \frac{3}{8} \, \ln (3 \Gamma) + \frac{\gamma}{2} - \frac{1}{3} \right] \; , \end{equation} 
\begin{equation} u_{ex}^{OCP} = - 0.898004 \, \Gamma + 0.96786 \, \Gamma^{1/4} 
      + 0.220703 \, \Gamma^{- 1/4} - 0.86097 \; . \end{equation}
$u_{ex}^{Abe}$ was derived by Abe \cite{ab.1} exactly in the giant 
cluster-expansion theory. $\gamma = 0.57721...$ is Euler's constant.   
$u_{ex}^{OCP}$ was evaluated by computer simulation of one 
component plasma (OCP), 
where a single species of charged particles embedded in a uniform 
background of neutralizing charges. 
$u_{ex} ( \Gamma )$ is derived for strongly coupled Coulombic plasma 
and we believe that it should be valid for any Coulombic plasma except a 
change, $\alpha \rightarrow 4 \alpha_s /3$ in $\Gamma$, for color Coulombic plasma. 
Finally, we get, 
\begin{equation} e(\Gamma) \equiv \frac{\varepsilon}{\varepsilon_s} = 1 + \frac{1}{3} u_{ex} (\Gamma) \; , 
\label{eq:e} \end{equation} 
where $\varepsilon_s \equiv 3 a_f T^4$, energy density of ideal QGP gas. 
Using the relation 
\begin{equation} \varepsilon = T \dy{P}{T} - P \; ,\end{equation} 
we get the pressure 
\begin{equation} p(T) \equiv \frac{P}{P_s} = \left( \frac{P_c}{a_f T_c} + 3 \int_{T_c}^T \, 
d\tau \tau^2 e(\tau) \right) / T^3 \; , \end{equation} 
where $P_s \equiv a_f T^4$ and $P_c$ is the pressure at critical temperature 
$T_c$. Temperature dependence of $e( \Gamma )$ seen in lattice data can only come 
from $\alpha_s$ in our model by using running coupling constant 
$\alpha_s (T)$. Let us use the running coupling constant $\alpha_s^L (T)$   
used in the lattice calculation \cite{bo.1} which may be obtained from 
the modified second order scaling relation, 
\begin{equation} \frac{\Lambda_L}{N_\tau T \lambda (\beta)} = ( 8 \pi ^2 \beta / 33)^{51/121} 
        \exp (- 4 \pi ^2 \beta /33) \; ,\end{equation} 
where $\beta \equiv 6/g^2 $ and $\Lambda_L, N_\tau$ are lattice parameters which may  
be replaced in terms of $T_c$ in above equation to get, 
\begin{equation} \frac{T}{T_c} = (\beta_c / \beta )^{51/121} 
     \exp (- 4 \pi ^2 (\beta_c - \beta) /33) \; ,\end{equation} 
where we have taken $\lambda (\beta) \approx \lambda (\beta_c)$, for simplicity. 
Also it follows from Ref. \cite{bo.1} that $\lambda (\beta)$ varies slowly as 
a function of $\beta$. After some algebra, we get, 
\begin{equation} \alpha_s^L (t) = \frac{2 \pi}{11 \left( \ln (t/t_0) +  
\frac{51}{121} \ln ( 12 \pi/33 \alpha_s^L (t)) \right)} \; , \label{eq:lsl} \end{equation}
where $t \equiv T/ T_c$ and $t_0$ is  
\begin{equation} t_0 = ( 8 \pi ^2 \beta_c / 33)^{51/121} 
        \exp (- 4 \pi ^2 \beta_c /33) \; . \end{equation} 
$\alpha_s^L (t) $ for given $t$ is obtained by solving Eq. (\ref{eq:lsl}). 
 $t_0$ or $\beta_c$ is only one parameter and adjusted to get the best 
fitting to lattice data \cite{bo.1}. 
In Fig. 1, we plotted $e(t)$, Eq. (\ref{eq:e}) and $p(t)$, 
\begin{equation} p(t) \equiv \frac{P}{P_s} = \left( p_c + 3 \int_1^t \, d\tau \tau^2 e(\tau) \right) / 
      t^3 \; , \end{equation} 
and obtained a remarkable 
good fit to lattice data \cite{bo.1} for $t_0 = .5763$ or 
$\beta_c = .57$. We have used $a_f = 16\; \pi ^2 /90$ which is for gluons. 
However, the value of $\beta_c$ is an order of magnitude 
smaller than that of lattice results. 
For small value of $\alpha_s$, $\alpha_s^L (t)$ 
may be approximated as    
\begin{equation} \alpha_s^L (t) \rightarrow \frac{2 \pi}{11 \ln (t/t_0)}  \left( 1 - 
\frac{51}{121} \frac{\ln (2 \ln (t/t0))}{\ln (t/t0)} \right)  \;, \end{equation}
or 
\begin{equation}  \alpha_s^L (t) \rightarrow \frac{2 \pi}{11 \ln (t/t_0)} \;. \end{equation}
The last expression is similar to the running coupling constant 
used in Ref. \cite{ri.1} where $t_0 \equiv \Lambda / T_c$ and also in 
Ref. \cite{bi.1} where $t_0 \equiv \Lambda / \pi T_c$. $\Lambda$ is the QCD scale 
parameter. For $\Lambda = 200 \;MeV$ we get $T_c = 347\; MeV$ and $T_c = 110\;MeV$ 
respectively. We have also plotted 
\begin{equation} \Delta = \frac{\varepsilon - 3 P}{3 a_f T^4} \; , \end{equation} 
a measure of nonidealness, and $c_s^2$. A reasonable good fit to 
lattice data is obtained, as shown in Fig. 2, except very close to $T_c$.  
In Fig. 3, we plotted $\alpha_s^L (t)$ and $\Gamma (t)$ as a function of $t$.  
We see that even at $T = 5 T_c$, gluon plasma is strongly coupled plasma 
and running coupling constant is not small.  

It is interesting to compare earlier few theories with our present 
theory. In Fig. 4, the best fits for $p(t)$ for three models
(Ref. \cite{pe.1}, Ref. \cite{ba.1} and present model)  are plotted. 
As we see the fitting of lattice data with the theory improves as we go from 
quasi-particle theory, our earlier result and present result. 
 
In conclusion, we obtained a remarkably good fit to the recent 
lattice result \cite{bo.1}  
using EOS of strongly coupled plasma of QED \cite{ic.1} with a modification for 
color degrees of freedom and with a running coupling constant which 
was used in lattice calculations. Hence we conclude that QGP is a 
strongly coupled Coulombic plasma of color charges. Just like in 
SCP, SCQGP may have a dramatic effects on various signatures, 
collective effects and other properties. They need to be recalculated 
before we make confirmation of QGP in relativistic heavy ion collisions. 
For e.g., screening length decreases rapidly near critical temperature 
and we may have serious consequences in $J/ \psi$ suppression results. 

I thank P. K. Kaw and J. C. Parikh for useful discussions on strongly 
coupled plasma. 

\begin{figure}
\caption { Plots of $p(t) \equiv P/P_s$ (dashed curve) and 
$e(t) \equiv \varepsilon / \varepsilon_s$ (continuous curve) as a function of $t \equiv T/T_c$ from 
our model and lattice results (dots), respectively. } 
\label{fig 1}
\vspace{.75cm}

\caption { Plots of $\Delta \equiv (\varepsilon - 3 P) / \varepsilon_s$ (continuous curve) and 
$c_s^2$ (dashed curve) as a function of $t \equiv T/T_c$ from 
our model and lattice results (dots), respectively. } 
\label{fig 2}
\vspace{.75cm}

\caption {Plots of $\alpha_s^L (t)$ (dashed curve) and $\Gamma (t)$ 
(continuous curve) as a function of $t$.}
\label{fig 3}
\vspace{.75cm}

\caption {Plots of the best fits for $p(t)$ from three models, 
Ref. [3] (dotted curve), Ref. [4] (dashed curve),   
present model (continuous curve) and lattice results (dots) 
as a function of $t$.} 
\label{fig 4}
\vspace{.75cm}

\end{figure}

\begin{thebibliography}{99} 
\bibitem{bo.1} G. Boyd, J. Engels, F. Karsch, E. Laermann, C. Legeland, 
M. Lutgemeier and B. Petersson, Phys. Rev. Lett. {\bf 75}, 4169 (1995); 
Nucl. Phys. {\bf B469}, 419 (1996).   
\bibitem{ri.1} D. H. Rischke, M. I. Gorenstein, A. Schafer, H. Stocker and 
W. Greiner  Phys. Lett. {\bf B278}, 19 (1992); Z. Phys. {\bf C56} 325 (1992).  
\bibitem{pe.1} A. Peshier, B. Kampfer, O. P. Pavlenko and 
G. Soff, Phys. Lett. {\bf B337}, 235 (1994). 
\bibitem{ba.1} V. M. Bannur, Phys. Lett. {\bf B362}, 7 (1995). 
\bibitem{en.1} J. Engels, J. Fingberg, F. Karsch, D. Miller and 
M. Weber, Phys. Lett. {\bf B252}, 625 (1990). 
\bibitem{ei.1} E. Eichten, K. Gottfried, T. Kinoshita, K. D. Lane and 
T. M. Yan, Phys. Rev. {\bf D17}, 3090 (1978). 
\bibitem{ic.1}  S. Ichimaru, {\it Statistical Plasma Physics (Vol. II) - 
Condensed Plasma} (Addison-Wesley Publishing Company, New York, 1994). 
\bibitem{ab.1} R. Abe, Progr. Theor. Phys. {\bf 21}, 475 (1959).
\bibitem{bi.1} T. S. Biro, C. Gong, B. Muller and A. Trayanov, 
Int. J. Mod. Phys. {\bf C5}, 113 (1994). 
\end{thebibliography}
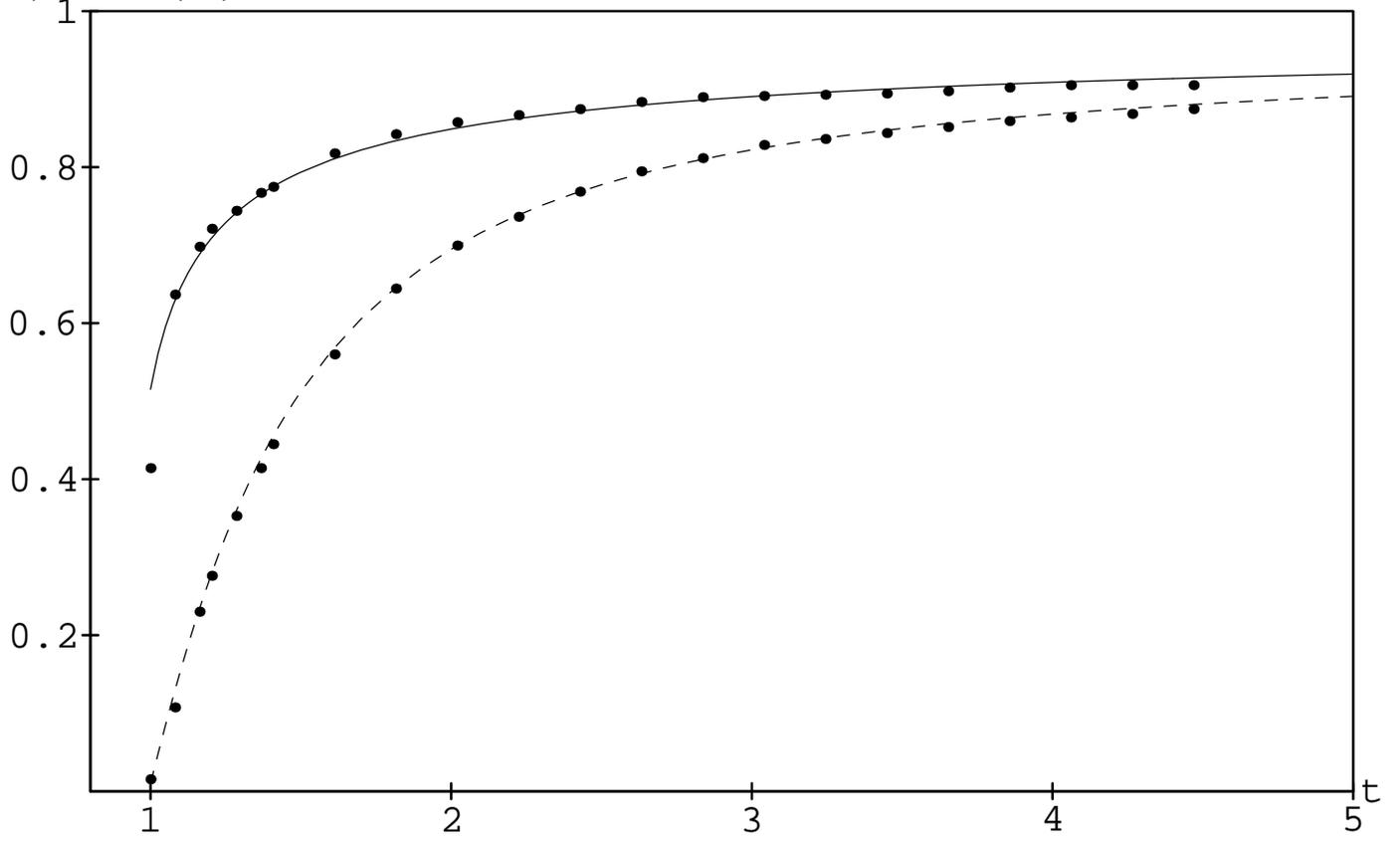
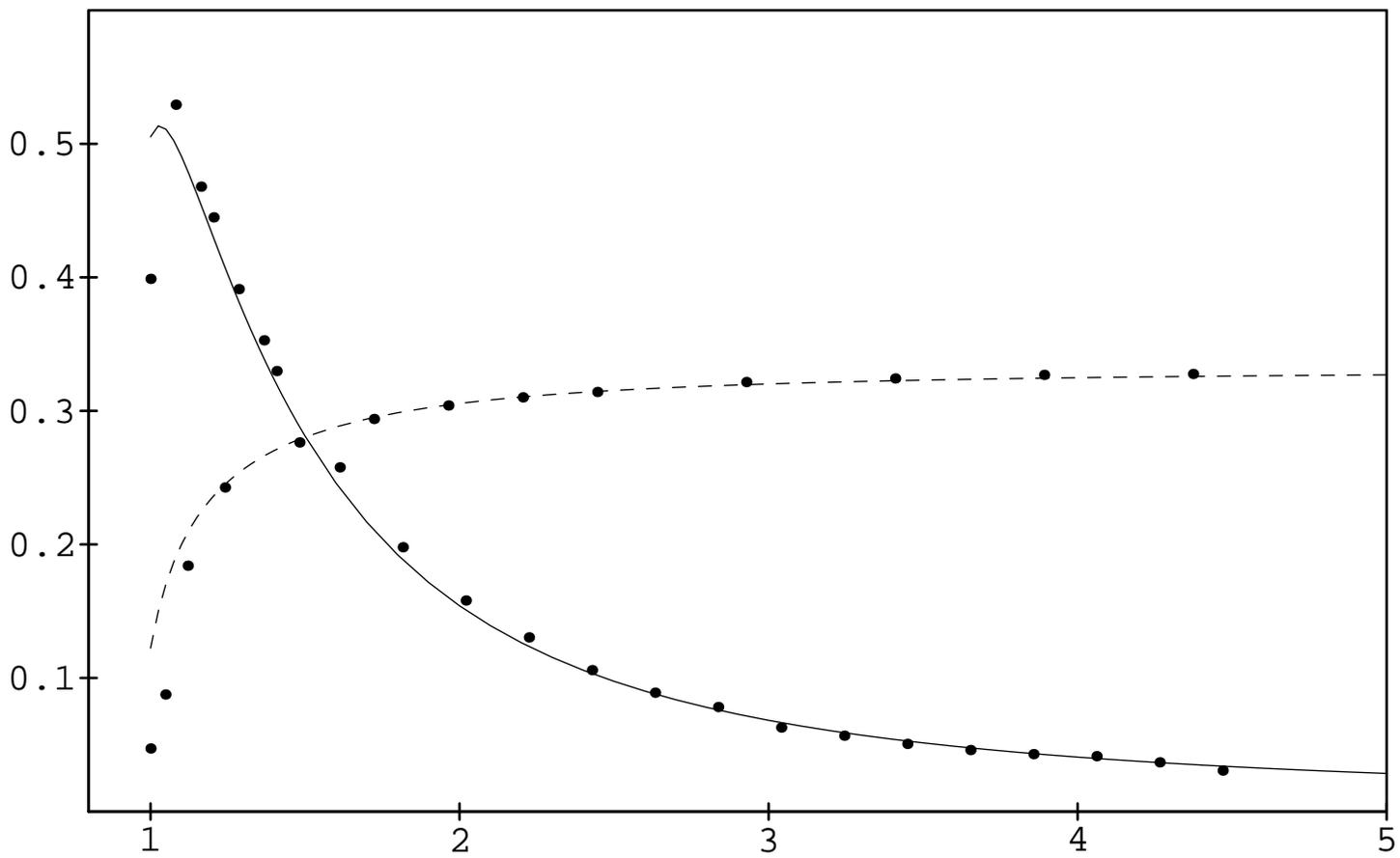
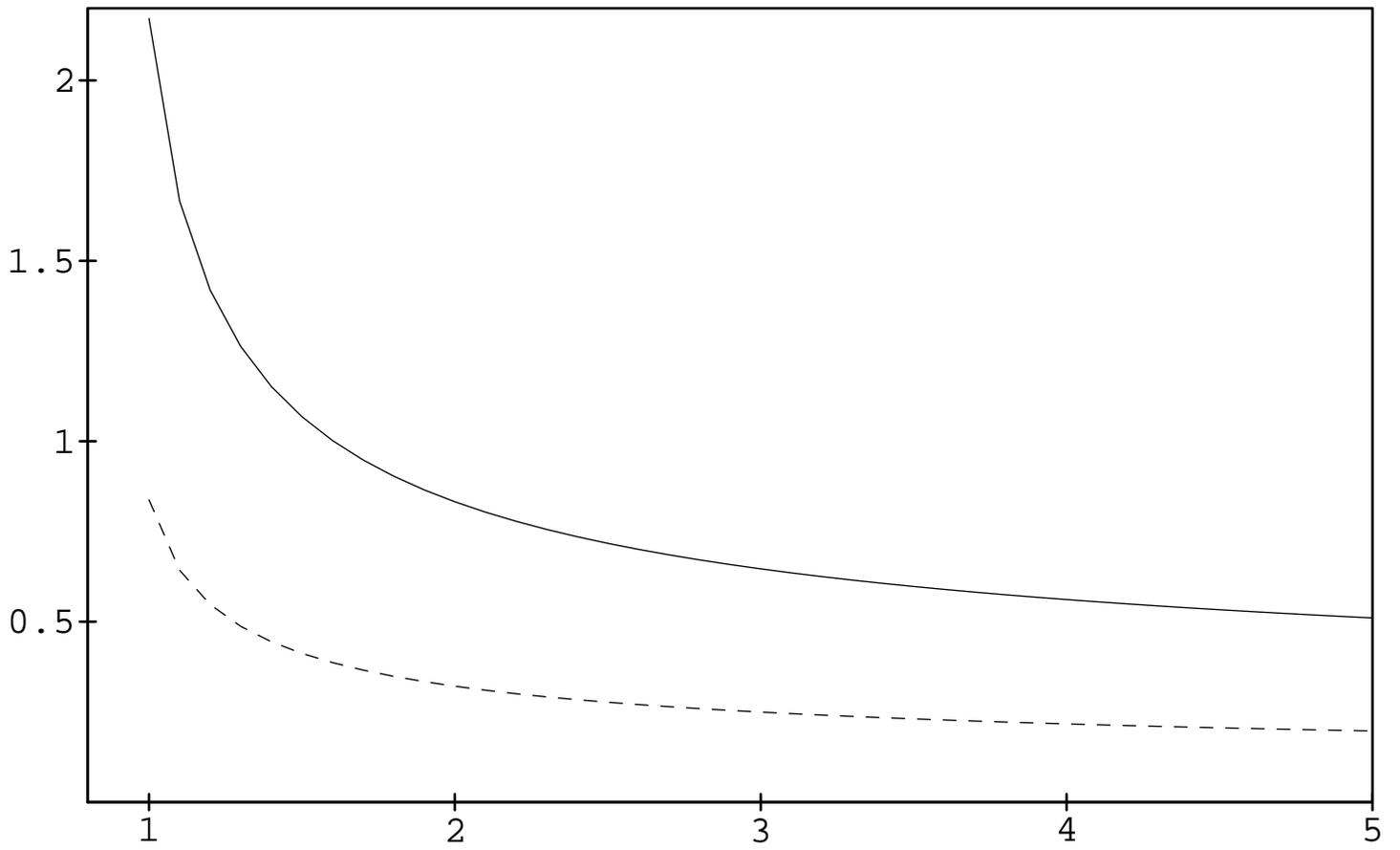
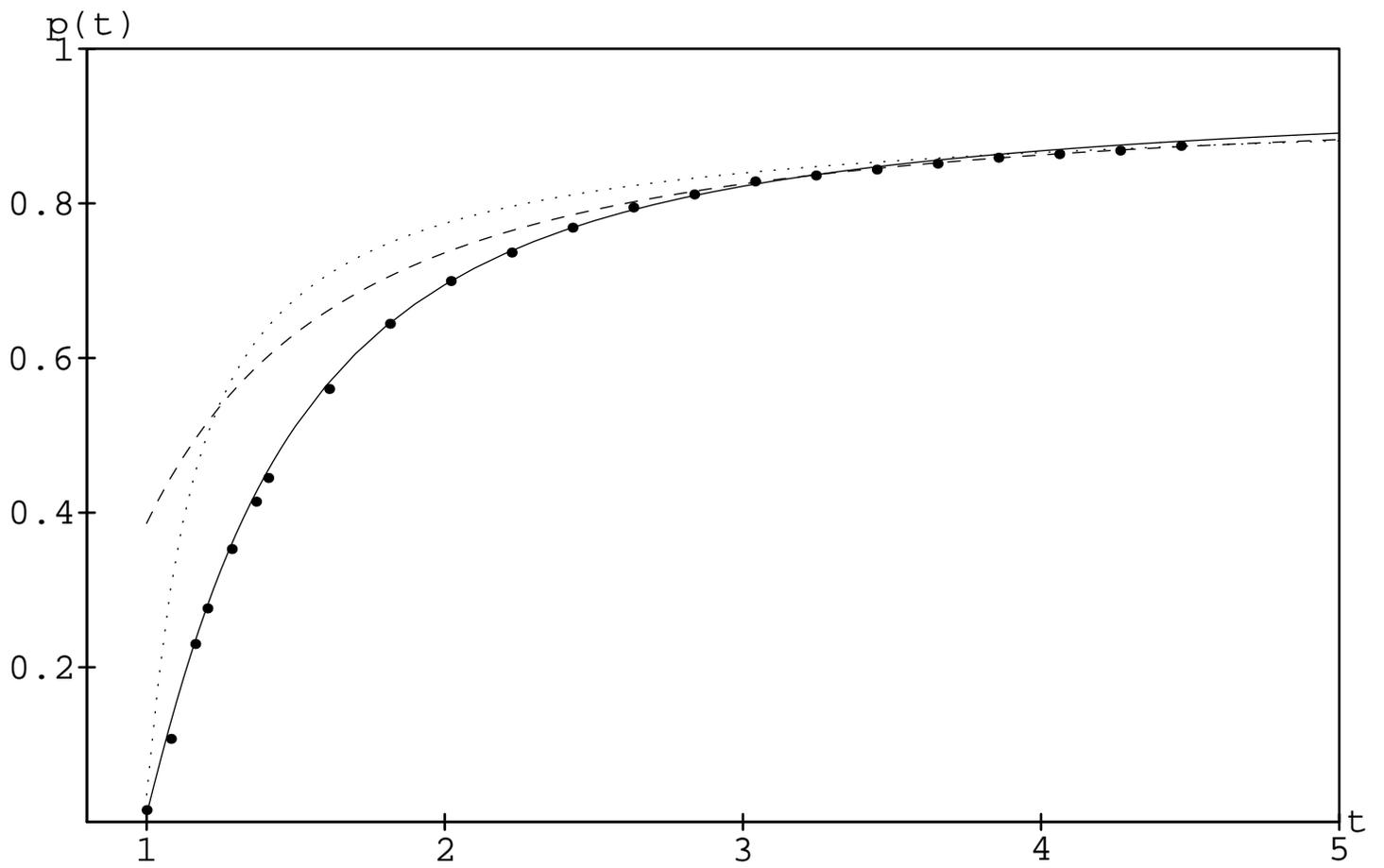
\end{document}